\documentclass{article}
\usepackage{graphicx} 

\usepackage{booktabs}   
\usepackage{caption}    
\usepackage{siunitx}    
\usepackage{amssymb}

\title{QOC DAO - Stepwise Development Towards an AI Driven Decentralized Autonomous Organization}
\author{Marc Jansen\textsuperscript{1,2}, Christophe Verdot\\University of Applied Sciences Ruhr West, BlockAI Ltd.\\marc.jansen@hs-ruhrwest.de, christophe@blockai.dev}
\date{}

\begin{document}

\maketitle
\begin{abstract}
This paper introduces a structured approach to improving decision-making in Decentralized Autonomous Organizations (DAOs) through the integration of the Question–Option–Criteria (QOC) model and AI agents. We outline a stepwise governance framework that evolves from human-led evaluations to fully autonomous, AI-driven processes. By decomposing decisions into weighted, criterion-based evaluations, the QOC model enhances transparency, fairness, and explainability in DAO voting. We demonstrate how large language models (LLMs) and stake\-holder-aligned AI agents can support or automate evaluations, while statistical safeguards help detect manipulation. The proposed framework lays the foundation for scalable and trustworthy governance in the Web3 ecosystem.
\end{abstract}

\section{Introduction}
As the Web3 ecosystem continues to mature, Decentralized Autonomous Organizations (DAOs) have emerged as a central governance model for blockchain-native communities, protocols, and ecosystems. By design, DAOs aim to eliminate centralized control and empower stakeholders to make collective decisions in a transparent and permissionless manner. However, in practice, the promise of decentralization often clashes with limitations in current governance processes. Empirical studies reveal persistent problems, including voting power imbalances, opaque decision rationales, low participation rates, and misaligned incentives. These issues not only undermine the fairness and legitimacy of DAO decisions but also hinder their capacity to make informed, long-term strategic choices.

At the same time, advances in Large Language Models (LLMs) and AI agents open up new opportunities to address these shortcomings. Recent developments in natural language reasoning, automated evaluation, and explainable AI suggest that human-AI hybrid systems could enhance governance by providing structured decision support, summarizing complex proposals, evaluating options against multi-dimensional criteria, and generating transparent rationales. However, to effectively integrate such systems into decentralized governance, we must first ground them in a formal decision-making methodology that is both human-interpretable and computationally tractable.

In this paper, we propose a structured framework for AI-augmented DAO governance based on the Question–Option–Criteria (QOC) model. Originally developed in the field of design rationale and human-computer interaction, the QOC framework decomposes decisions into three fundamental components: (1) the central question to be answered, (2) the set of possible options, and (3) the evaluation criteria that reflect community priorities. By assigning weights to each criterion and collecting structured evaluations for how well each option satisfies these criteria, the model enables both quantitative aggregation and qualitative insight into the rationale behind collective decisions.

We present a stepwise development path for integrating QOC into DAO governance. The first step involves a purely human-driven process in which DAO participants manually evaluate proposals using a predefined set of criteria and weights. The second step introduces AI agents that simulate stakeholder perspectives and generate evaluations, while humans retain final decision authority, a human-in-the-loop governance model. In the final step, we envision a fully autonomous, AI-driven DAO, in which stakeholder-aligned agents conduct the evaluation and decision, making process end-to-end, with comprehensive reporting and traceability.

Throughout this evolution, the QOC model serves as a unifying structure that supports transparency, fairness, and explainability. It not only facilitates the integration of LLMs and agent-based AI into governance workflows, but also mitigates the risks associated with token-weighted voting, such as collusion, voter apathy, and manipulation. Additionally, the QOC framework provides the foundation for statistical safeguards, such as outlier detection, that enhance robustness against biased or malicious voting behavior.

By embedding structured reasoning and AI augmentation into the heart of DAO decision-making, our approach seeks to move beyond simplistic yes/no voting toward a more deliberative, auditable, and intelligent governance paradigm. The remainder of this paper is structured as follows: Section 2 reviews the current state of DAO governance and its limitations. Section 3 introduces the QOC model in detail and outlines its mathematical formulation. Section 4 presents the stepwise implementation path toward AI-driven DAOs, and Section 5 discusses opportunities for empirical evaluation. Finally, Section 6 outlines future research directions and open challenges in building trustworthy, AI-augmented governance systems for the Web3 era.

\section{State of the Art}
The governance of DAOs has emerged as a focal point of both innovation and critique within the Web3 ecosystem. While DAOs aspire to embody decentralized, transparent, and community-led decision-making, recent research highlights persistent structural and procedural challenges that hinder these ideals. This section reviews the current state of DAO governance, focusing on well-documented issues such as voting power imbalances, lack of transparency in decision rationales, and low community engagement. It then presents emerging solutions, including structured decision frameworks and AI-assisted governance models, that aim to improve both the legitimacy and effectiveness of decentralized decision-making.
\subsection{Challenges in Current DAO Governance}
Decentralized Autonomous Organizations promise transparent, community-driven governance, but in practice they face several well-documented challenges. Recent studies highlight at least three core issues in the typical token-based voting model \cite{intelligent_daos}:
\begin{itemize}
    \item Concentration of voting power (the “whale problem”): A small number of large token holders often control a disproportionate share of votes. For example, analyses show that in many DAOs fewer than ten actors hold over half of the voting power \cite{dao_design}. This concentrated ownership (colloquially, whales) creates a unique agency problem, the interests of whales may diverge from those of smaller token holders, leading to governance vulnerabilities \cite{dao_review}. When some powerful members can unilaterally influence decisions, it undermines the democratic ethos of DAOs and can discourage broader participation\cite{blockai_medium}.
    \item Low voter participation and engagement: Despite open governance ideals, most DAOs struggle with voter apathy. Empirical data indicate that turnout is often very low, even for important proposals. For instance, one study noted DAO-wide participation rates as low as 20–30\% of token holders on critical votes \cite{intelligent_daos}. Such low engagement means decisions can be made by a small minority, further amplifying the influence of whales and calling into question the legitimacy of “community” decisions \cite{dao_design}. The complexity of proposals and the cognitive burden on voters are cited as factors for disengagement – many members simply do not feel informed enough to vote \cite{intelligent_daos}.
    \item Misaligned incentives and short-termism: In token-based governance, short-term speculators and long-term stakeholders may have conflicting goals. Research points out that speculative behavior (seeking immediate token gains) can clash with decisions aimed at long-term protocol sustainability \cite{intelligent_daos}. This misalignment can fragment the community’s priorities and reduce coherent engagement in governance. Participants focused on quick profits might abstain from votes about long-term investments or upgrades, leaving such decisions to a few committed members. Over time, these structural issues contribute to inefficient decision-making and erosion of trust in the DAO’s process \cite{intelligent_daos}.
\end{itemize}
Another practical challenge is how voting is conducted. Many DAOs use off-chain voting platforms (such as Snapshot) for convenience, but this introduces transparency and security trade-offs. Off-chain voting lacks the full immutability and auditability of on-chain processes: votes are not permanently recorded on the blockchain, which can lead to trust issues and challenges in verifying outcomes. An empirical analysis by Bellavitis and Momtaz (2024) found that DAOs using non-algorithmic off-chain voting suffered a significant discount in their valuation, suggesting that deviations from decentralized, transparent governance are costly. In contrast, on-chain voting aligns with blockchain’s principles of transparency and accountability, providing an immutable public record of decisions. However, on-chain voting can be resource-intensive (due to gas fees and technical overhead), so many organizations opt for off-chain votes at the expense of some transparency. This trade-off between efficiency and openness poses a governance dilemma \cite{dao_votings}.

\subsection{Lack of Transparency and Rationale in Binary Decisions}
Most DAO governance today boils down to binary decisions on proposals – essentially “yes” or “no” votes on a single option\cite{blockai_medium}. Once the vote concludes, the community typically only sees the outcome tallies (e.g. 60\% yes, 40\% no) and whether the proposal passed or failed. There is usually no deeper insight into why a decision was made. As a recent analysis emphasizes, after a binary vote “all the community sees are the final tallies,” with no record of the reasoning behind votes or the criteria that mattered to participants. In other words, the DAO knows what was decided but not why. This opacity can make decisions feel arbitrary or unaccountable, especially to members who disagreed with the outcome.

The lack of structured feedback means that valuable information is lost. For example, if a funding proposal is rejected, the binary result does not reveal whether voters opposed it due to concerns about the project’s concept, its execution plan, budgeting issues, or timing. Without this knowledge, proposal creators struggle to address the community’s actual objections – they receive a blunt “no” without context. This not only frustrates proposers but also means the DAO forgoes the opportunity to learn and iterate on ideas. The decision-making process thus remains a “black box” to the wider community, reducing trust in governance outcomes\cite{blockai_medium}.

Researchers have pointed out that this binary, uncontextualized voting approach hampers transparency and accountability in DAOs\cite{blockai_medium}. When decisions are made without explicitly stated reasons, it becomes difficult for members (especially those on the losing side of a vote) to accept and support the outcome. In contrast, traditional organizations often accompany decisions with minutes, reports, or rationale which, while centralized, at least document why a choice was made. DAOs currently lack a native mechanism for recording decision rationales on-chain. As a result, much of the real discussion and deliberation happens off-chain (e.g. in forums or Discord), and only the final yes/no result is captured on-chain. This separation further diminishes transparency, outsiders or later participants cannot easily trace the arguments or criteria considered, unless they manually sift through discussion threads.

Overall, the literature suggests that merely opening governance to token holders does not automatically yield transparent or merit-based decisions. Without frameworks to capture the criteria and arguments behind votes, DAO governance can devolve into popularity contests or be swayed by informal, off-chain persuasion that escapes public scrutiny. Transparency in a DAO should ideally extend beyond seeing the vote totals to understanding the decision-making process itself\cite{blockai_medium}. This gap in current implementations has motivated researchers and practitioners to seek more structured decision-making models for DAOs.

\subsection{Structured Decision Frameworks for DAOs}
To address the above problems, recent work has proposed introducing structured decision-making frameworks into DAO governance. One such approach is the Question–Option–Criteria (QOC) model, a methodology originally developed in design rationale research to systematically evaluate alternatives against explicit criteria. In the context of DAOs, a QOC framework would reshape a proposal decision into a well-defined question (e.g., “Should we fund project X?”), a set of options (e.g., Yes or No – or potentially multiple alternative proposals), and a set of predefined criteria on which to evaluate those options. Participants would rate how well each option satisfies each criterion, and these evaluations would be aggregated (often with weighting of criteria) to determine the decision outcome\cite{blockai_medium}.

The appeal of QOC in a DAO is its potential to make the decision process more transparent, reasoned, and resistant to manipulation. Instead of a simple yes/no tally, the community would produce a decision profile that shows the scores of each option on each criterion and reveals which factors drove the decision. For example, a funding proposal might be evaluated on criteria such as expected ROI, alignment with the DAO’s mission, technical feasibility, and security risk. If the proposal is rejected, the QOC results might show that it scored well on ROI but poorly on security and mission alignment, providing information far more actionable than a flat rejection. By systematically recording the options considered, the criteria applied, and the evaluations made, the QOC approach inherently documents the rationale behind decisions\cite{blockai_medium}. This creates an audit trail of governance reasoning, which improves accountability and allows the community to revisit and learn from past decisions.

Notably, structured multi-criteria decision methods have proven useful in other domains for improving group decisions. In software engineering and business management, Multi-Criteria Decision-Making (MCDM) frameworks and Decision Support Systems have been used to evaluate alternatives against weighted criteria in a rigorous way\cite{decision_models}. These systems often include discussion and negotiation features to help stakeholders reach consensus based on evidence and priorities. Applying similar principles to DAO governance could help elevate decisions from pure token-weighted polls to merit-based choices informed by data and agreed-upon values. Indeed, one recent study on community grant allocation (an inherently subjective process) found that adopting DAO-inspired voting mechanisms improved perceived fairness and inclusivity in decision-making\cite{dao_design}. By borrowing concepts like weighted criteria and community evaluation from DAO governance, the process became more impartial – reducing bias and “personal network” influence that had plagued the traditional method\cite{dao_design}. This example underscores that innovative voting and deliberation models from the blockchain world can have broader benefits for transparency and fairness in group decisions.

Beyond QOC, the DAO space has seen experiments with alternative voting mechanisms to overcome the limitations of simple majority voting. For instance, quadratic voting has been tested in some DAOs as a way to curb whale dominance – it makes the cost of votes quadratic, so that casting multiple votes becomes exponentially more expensive for large holders, thereby diminishing their relative influence\cite{dao_voting_mechanisms}. Another innovation is conviction voting, where votes accumulate over time to continuously measure community support, rather than a one-time yes/no snapshot\cite{dao_design}. Additionally, vote-escrow (lockup) models (as seen in protocols like Curve) require voters to lock tokens for a period to weight their voting power, aligning influence with long-term commitment. Each of these mechanisms attempts to address fairness, nuance, or commitment in decision-making. For example, Tamai et al. (2024) found that quadratic voting alone, while mitigating simple whale domination, is still vulnerable to collusion (coordinated groups exploiting the system)\cite{dao_voting_mechanisms}. They proposed a hybrid model combining quadratic voting with vote-escrowed tokens, which their simulations showed could reduce both the whale problem and collusion risk in DAO voting. This indicates that no single tweak is a silver bullet, a blend of mechanisms and careful design is needed to balance decentralization with decision quality.

The QOC approach complements these innovations by tackling the information side of governance rather than purely the voting power distribution. It ensures that, whatever voting mechanism is used, the decision is based on explicit evaluation of criteria. Importantly, by capturing the community’s reasoning, QOC could also deter manipulative practices. If a whale’s large vote goes against the trend of evaluated criteria, it would be evident in the record, potentially prompting scrutiny or justification. In essence, structured frameworks like QOC bring an element of deliberative democracy into the DAO: decisions are not just votes, but reasoned arguments. This can enhance legitimacy – members are more likely to accept outcomes when they see a fair evaluation process – and it can improve decisions themselves, as decisions made with reference to well-defined criteria tend to be higher quality and more consistent\cite{decision_models}.

\subsection{Integrating AI into DAO Decision-Making}
Looking forward, scholars and practitioners are considering the role of Artificial Intelligence in addressing DAO governance challenges. One emerging vision is a phased integration of AI into DAO governance, moving from human-driven decisions to partially, and eventually fully, AI-driven processes\cite{blockai_medium}. This trajectory aligns with a general “human-in-the-loop” approach popular in AI: initially use AI to assist and augment human decision-makers, and gradually increase automation as trust in the AI grows.

In the near term, AI can serve as a decision-support tool for DAOs. A recent case study by Chen et al. proposes an AI-assisted governance framework to help overcome voter information overload and low participation\cite{intelligent_daos}. The framework uses large language models (LLMs) with chain-of-thought (CoT) reasoning to analyze proposals and generate structured, explainable recommendations for voters. In practical terms, the AI might summarize a complex proposal, evaluate its potential impacts (e.g. on the DAO’s treasury or tokenomics), and even map it against criteria or stakeholder viewpoints. By providing clear insights and highlighting the pros/cons of a proposal, the AI lowers the cognitive barrier for members to participate. The study reported promising results: the AI’s recommendations had about 97\% alignment with historical human-made decisions (indicating it mirrored the community’s likely judgment), and simulations suggested that such AI assistance could increase voter turnout by around 40\% and improve perceived governance clarity by 35\%. In other words, an AI decision aide can help make governance more inclusive (more voters willing to engage) and more transparent (decisions come with a machine-generated rationale or analysis).

Crucially, these AI systems are designed to be explainable. Techniques like chain-of-thought ensure that the AI not only provides a recommendation but also the reasoning path behind it\cite{intelligent_daos}. This focus on explainability is essential for building trust in AI-driven governance, in which DAO members need to understand why the AI is suggesting a certain action, much like they need to understand each other’s arguments in a deliberation. By making AI’s reasoning transparent, we preserve accountability even as we introduce automation.

The human-in-the-loop model for DAOs would work as follows: AI evaluates each option against the agreed criteria (much as humans would do in the QOC approach) and perhaps even suggests an outcome, but the community retains the final say to accept or override the AI’s suggestion\cite{intelligent_daos}. This approach leverages the efficiency and analytical power of AI while keeping human judgment at the helm for value-laden decisions. It is analogous to an AI-recommendation system in corporate boards or governments, where algorithms might analyze data and present policy options, but elected humans make the ultimate choice. Such a model has been described as a hybrid governance model, marrying automated analysis with human oversight.

The literature suggests that over time, if the AI proves reliable and the community gains confidence in its impartiality, the balance could shift more toward automation. Some blockchain thinkers even envision fully AI-driven DAOs in the future, organizations where intelligent agents, coded with the community’s goals and constraints, autonomously make funding decisions, investment choices, or policy updates\cite{blockai_medium}. In this final stage, human involvement would be minimal, perhaps only to set high-level objectives or intervene if the AI behaves undesirably. The potential advantages of AI-driven DAOs include speed, consistency, and the removal of human biases or collusion. However, research also cautions that reaching this stage requires extreme transparency and trust in AI that is not yet achieved. Ensuring that an AI’s objectives remain aligned with the community (avoiding “model misalignment” where an AI might optimize the wrong metric) is a major challenge highlighted in discussions about DAO 2.0 and AI governance\footnote{https://www.bankless.com/read/rise-of-ai-driven-daos\#:~:text=The\%20Rise\%20of\%20AI,\\DAO\%20optimizes\%20for\%20metrics} \footnote{https://www.panewslab.com/en/articles/hllcge5w}. In other words, who governs the governors becomes what governs the algorithm.

In summary, recent academic literature and experimental implementations are converging on a vision of more intelligent, accountable DAO governance. First, address the current pain points: the simplistic yes/no voting schemes, lack of rationale transparency, whale-dominated outcomes, and voter apathy. Next, introduce structured decision methodologies (like QOC and multi-criteria evaluation) to inject transparency, deliberation, and fairness into the process. Finally, augment and possibly automate this process with AI – initially as a supportive tool that provides data-driven analyses and recommendations (the human-in-the-loop paradigm) and eventually as a principal decision-maker once it earns sufficient trust and oversight mechanisms are in place. The end goal is a DAO that not only embodies decentralized and democratic ideals in theory, but also achieves effective, explainable, and inclusive governance in practice\cite{intelligent_daos}.

Throughout this evolution, prioritizing peer-reviewed, transparent approaches is key. Many of the proposals (from quadratic voting to AI-co-pilots for governance) are still in experimental stages, being refined through academic research and pilot projects. What is clear from the literature of the past five years is that “naïve” DAO governance, simply token voting on yes/no questions, has significant shortcomings\cite{blockai_medium}. The research community is actively exploring remedies, combining insights from computer science, economics, and political science. By learning from these studies and trials, we can design DAO governance models that are both decentralized and well-reasoned, capturing not just the voice of the majority, but the nuanced rationale of the community in each decision. Such advancements are expected to improve the legitimacy and effectiveness of DAOs, paving the way for more autonomous yet accountable organizations in the Web3 ecosystem.

\section{The Question-Option-Criteria (QOC) Model for DAO's}
This section introduces the QOC model as a structured framework for transparent and reasoned decision-making. It begins with a general overview of the model’s core components, questions, options, criteria, weights, and evaluation functions, to establish the foundation for later adaptations. Building on this, we explore how the model can be tailored to the specific needs of DAOs, including multi-user input, binary voting, and mechanisms to enhance transparency and fairness.

\subsection{The QOC Model}
The QOC (Question–Option–Criteria)\cite{qoc} framework is a structured decision-making method designed to facilitate transparent, reasoned evaluation processes. It breaks down complex decisions into three core elements, Questions, Options, and Criteria, and incorporates quantitative mechanisms for comparative analysis.
\subsubsection{O - The Question}
At the heart of the framework lies the Question (Q), which defines the central decision-making problem. This question should be:
\begin{itemize}
    \item Well-formulated: Clearly articulated, concise, and unambiguous.
    \item Actionable: Framed in a way that it can be answered through a selection from a set of defined options.
    \item Decisive: Typically phrased to invite a single best answer (e.g., “Should we fund project X?” or “Which algorithm is most suitable for our use case?”).
\end{itemize}
\subsubsection{O - The Options}
The Options (O) represent the possible answers or solutions to the question posed. Formally, we define a set of options:
\\
\begin{center}
$O=(o_0, o_1, ..., o_n)$
\end{center}
Each option $o_k$ (with $k\in\{0,…,n\}$) is a discrete and independent alternatives, mutually exclusive and meant to address the decision question. In many practical use cases, such as the in this paper discussed DAO votes, the options might simply be binary (e.g., Yes or No), but the framework in general supports arbitrary cardinality.

\subsubsection{C - The Criteria}
The Criteria C are the standards or dimensions along which each option will be evaluated. Formally:
\\
\begin{center}
$C=(c_0, c_1, ..., c_m)$
\end{center}
Each criterion $c_j$ (with $j\in{0,…,m}$) reflects an aspect of quality, performance, or desirability that matters for the decision. Examples include cost, feasibility, security, user acceptance, or expected return on investment.

\subsubsection{W - The Weights of Criteria}
Each criterion $c_j$ is assigned a weight $w_j in (0, 100]$ indicating its relative importance in the decision process. These weights can be:
\begin{itemize}
    \item Normalized to sum up to 100, representing percentage influence.
    \item Ranked independently if strict normalization is not required.
\end{itemize}
The vector of weights is denoted:
\\
\begin{center}
$W=(w_0, w_1, ..., w_m)$
\end{center}
Higher weights reflect greater importance in the final evaluation.

\subsubsection{E - Evaluation Function}
The evaluation function $e(k,j)$ maps each Option–Criterion pair to a support score in the range $[0, 100]$. Formally:
\\
\begin{center}
$e : O \times C \rightarrow [0,100] \subset \mathbb{N}$
\end{center}
Each value $e(k,j)$ represents how well the option $o_k$ satisfies or supports the criterion $c_j$:
\begin{itemize}
    \item A value of 0 indicates no support at all.
    \item A value of 100 indicates maximal support or alignment.
    \item Intermediate values reflect partial or moderate support.
\end{itemize}
These scores can be generated by:
\begin{itemize}
    \item Human judgment (manual scoring by experts or stakeholders).
    \item Automated analysis (e.g., via AI or data-driven evaluation).
    \item Hybrid methods (e.g., AI-generated suggestions reviewed by humans).
\end{itemize}

\subsubsection{Final Scoring and Decision}
To determine the most suitable option, the overall score
$S(o_k)$ for each option $o_k$ is computed as a weighted sum of evaluation scores:
\\
\begin{center}
$S(o_k) = \sum_{j=0}^{m} w_j \cdot e(k, j)$
\end{center}
The option with the highest overall score is considered the most appropriate response to the original question $Q$, assuming all weights and evaluations are valid and trustworthy.

\subsection{A Potential Adaptation of the QOC Model for DAO's}
The following subsections describe potential modifications to the model, to make it usable in the context of DAO's.

\subsubsection{Extend the QOC Model to a Multi User Based Approach}
While the QOC model, as described in the previous section, already provides a structured and transparent approach to decision-making by collecting options, criteria, weights, and evaluations from a single user, its full potential unfolds when extended to a multi-user setting. In this enhanced variant, multiple stakeholders are involved in the process, each contributing their own perspectives by suggesting additional options and criteria relevant to the decision question. After a consolidation phase, in which redundancies and overlaps are resolved and the collective set of options and criteria is finalized, the same group of stakeholders proceeds to assign weights to the criteria and evaluate how well each option satisfies each criterion. This collaborative approach increases the richness and representativeness of the input data, fosters participatory decision-making, and ensures that the resulting evaluations reflect a more diverse and balanced perspective on the question at hand.\\
Let $U=(u_0, u_1, ..., u_r)$ be the set of users or stakeholders participating in the decision process. We could then adapt the above presented formulas as following:\\
\newline
Assuming that each user $u_i$ (with $i$ in ${1, ..., r}$) contributes their own individual set of options::
\\
\begin{center}
$O_i=(o_{i, 0}, o_{i, 1}, ..., o_{i, n})$
\end{center}
A consolidated set of options can be constructed by:
\\
\begin{center}
$O = \bigcup_{i=1}^{r} O_i = \{ o_0, o_1, \dots, o_n \}$
\end{center}
An analogous approach is applied to the criteria contributed by the individual users:
\\
\begin{center}
$C_i=(c_{i, 0}, c_{i, 1}, ..., c_{i, m})$
\end{center}
and the consolidated set of criteria, described as:
\\
\begin{center}
$C = \bigcup_{i=1}^{r} C_i = \{ c_0, c_1, \dots, c_m \}$
\end{center}
Additionally, each user $u_i$ provides a weight vector:
\\
\begin{center}
$W_{i} = \{ w_{i, 0}, w_{i, 1}, \dots, w_{i,m} \}$
\end{center}
which can be aggregated using the following formula:\\
\begin{center}
$\bar{w}_j = \frac{1}{r} \sum_{i=1}^{r} w_{i, j}$
\end{center}
An evaluation score is also assigned by each user $u_i$:
\\
\begin{center}
$e_i(k, j)$
\end{center}
The aggregated evaluation over all users is computed as follows:\\
\begin{center}
$\bar{e}(k, j) = \frac{1}{r} \sum_{i=1}^{r} e_i(k, j)$
\end{center}
The final score for each option $o_k$ is computed as follows:
\\
\begin{center}
$S(o_k) = \sum_{j=0}^{m} \bar{w}_j \cdot \bar{e}(k, j)$
\end{center}

\subsubsection{Adapt the QOC Model to the Typical Situation in DAO's}
In most Decentralized Autonomous Organizations (DAOs), decision-making revolves around straightforward, binary questions, typically of the form: "Should the DAO support initiative X?" This recurring structure provides an opportunity to streamline the QOC model by adapting it to a fixed-question, fixed-option scenario.

Instead of defining a new question and a new set of options for each vote, we can assume a standardized question template, such as "Should the DAO approve or support proposal X?" Correspondingly, the set of options is reduced to just two: Yes and No. This simplification significantly reduces the complexity of each voting process, while still enabling meaningful evaluations.

To further decrease decision-making overhead and to align all evaluations with the overarching goals of the DAO, it is also possible to predefine a fixed set of evaluation criteria. These criteria, such as return on investment, technical feasibility, alignment with the DAO’s mission, or risk exposure, can be determined once and then reused for all future decisions. If desired, the corresponding weights that reflect the relative importance of each criterion can also be defined upfront, either by consensus or based on the DAO’s governance rules.

With both the question, the options, and the weighted criteria fixed, the only remaining task for participants in each vote is to evaluate how well each of the two options (Yes or No) satisfies each of the predefined criteria in the context of the specific proposal. In other words, the focus shifts entirely to filling out the evaluation matrix, assigning scores that express how strongly each option supports each criterion.

This adaptation allows DAOs to leverage the structured, multi-criteria strength of the QOC model, while maintaining a user-friendly and scalable voting process. It ensures that decisions are not only democratic but also systematically aligned with long-term priorities, without requiring stakeholders to redefine the evaluation framework for every new proposal.

\subsubsection{Increase Transparency and Fairness in DAOs}
While the previous sections have demonstrated how the QOC model can be structurally adapted to the context of DAOs, its true value lies in what it enables: significantly enhanced transparency and fairness in decentralized governance. Unlike traditional DAO voting mechanisms, which typically yield only binary outcomes ("yes" or "no"), a QOC-based process records granular evaluations for each predefined criterion. This creates the opportunity to generate a detailed, criterion-level decision report after each vote, offering the community actionable insight rather than a black-box verdict.
\begin{figure}[htb]
    \centering
    \includegraphics[width=0.7\textwidth]{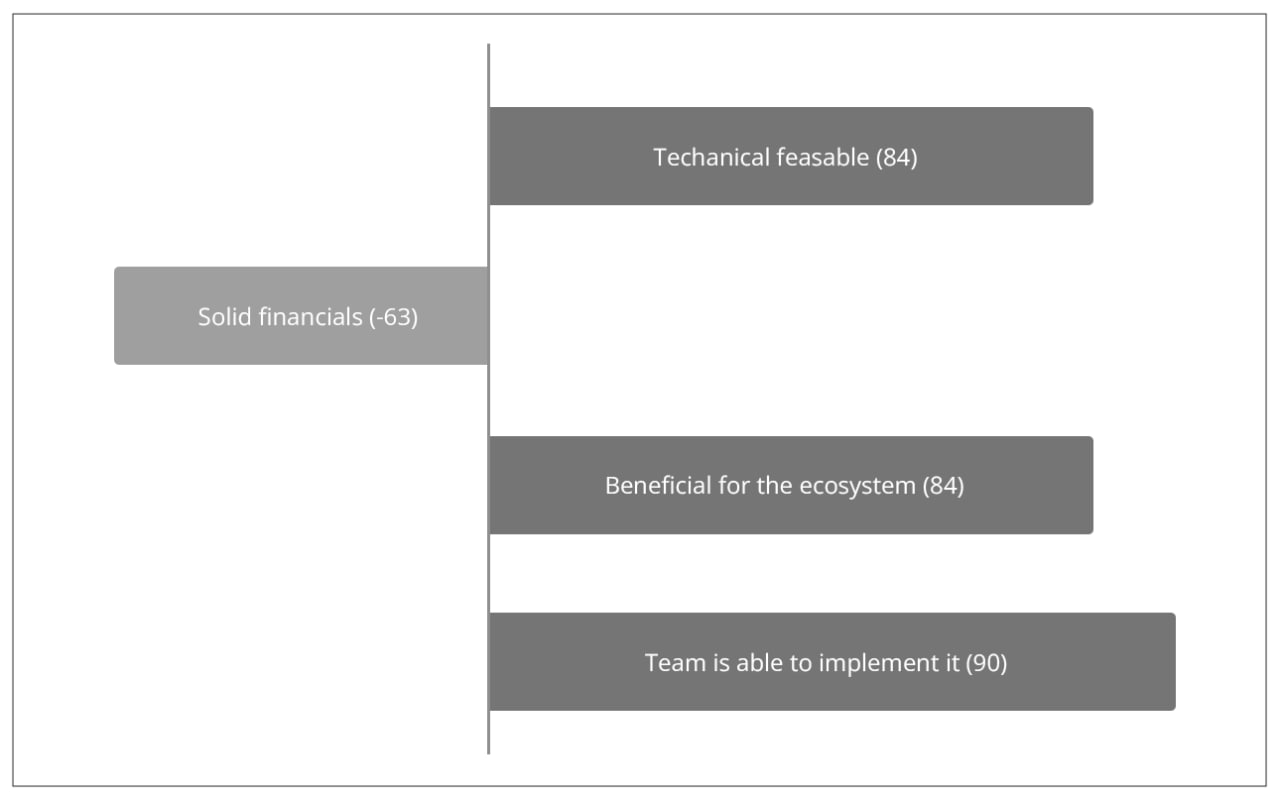}
    \caption{Example of criterion-wise evaluation in a DAO vote.}
    \label{fig:dao_vote}
\end{figure}
For instance (as shown in Figure \ref{fig:dao_vote}), a proposal may be evaluated positively with respect to "ecosystem benefit" and "technical feasibility," while receiving low scores on "financial sustainability". If the proposal is rejected, the proposers can use this multidimensional feedback to iteratively improve the proposal in alignment with community concerns before resubmitting it.

Beyond transparency, the QOC model also introduces mechanisms to improve fairness. In conventional token-weighted voting systems, outcomes can be disproportionately influenced by large stakeholders or coordinated voting blocs. While QOC-based voting is not immune to strategic behavior, such as participants deliberately misrepresenting scores across criteria, it provides a structured data layer that allows for the detection of anomalous behavior. By analyzing deviations between individual evaluations and aggregated community norms, the DAO can implement statistical safeguards against manipulation.
\begin{figure}[htb]
    \centering
    \includegraphics[width=0.7\textwidth]{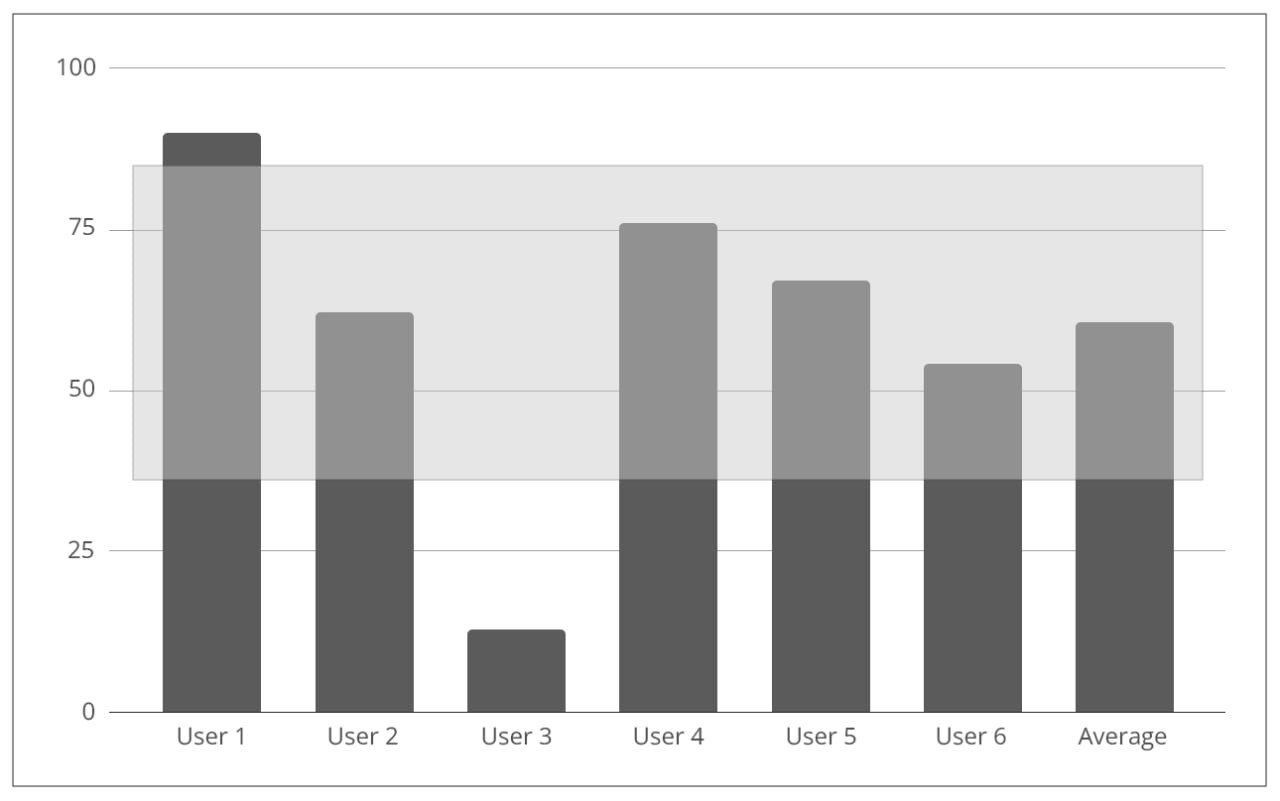}
    \caption{Example of outlier detection.}
    \label{fig:outlier_detection}
\end{figure}
One effective approach is outlier detection (see Figure \ref{fig:outlier_detection}, where individual criterion evaluations are compared to the community average using standard deviation thresholds. Evaluations that fall significantly outside the typical range can be flagged as outliers. The DAO may then choose to discount such outlier evaluations in the final aggregation process, thereby reducing the impact of extreme or intentionally biased input.

By combining transparent rationales for decisions with statistical protections against manipulation, this enhanced QOC-based voting model reshapes the nature of DAO governance. It moves the decision-making process away from simple vote tallies and toward a system rooted in community reasoning and defensible, criterion-based assessment. As a result, governance outcomes become more interpretable, more resistant to manipulation, and ultimately, more legitimate in the eyes of the participating community.

\section{Stepwise Implementation Towards a Fully AI Driven Solution}
To translate the conceptual QOC model into practical DAO governance, we propose a stepwise implementation path toward increasing levels of AI integration. This approach begins with a fully human-driven process, progresses to a hybrid human–AI model, and ultimately culminates in a fully autonomous, AI-driven governance system. The following subsections describe these stages in detail, illustrating how transparency, scalability, and fairness evolve at each step.

\subsection{Step 1: DAOs Based on The QOC Idea}
In the initial implementation phase, a DAO is deployed that adheres to the QOC-based governance model described above. In this first version, the set of evaluation criteria is predefined and globally fixed, with associated weights reflecting the strategic goals and values of the respective DAO. These criteria are intended to ensure consistent alignment across proposals and simplify the evaluation process for participants.

During each voting cycle, stakeholders are invited to assess how well the submitted project or initiative satisfies each of the established criteria. These assessments are expressed as numerical scores, and each participant’s input is weighted according to their respective voting power (e.g., based on token holdings or reputation scores). The resulting scores form the basis for computing the overall evaluation of each proposal.

\begin{figure}[htb]
\centering
\includegraphics[width=0.7\textwidth]{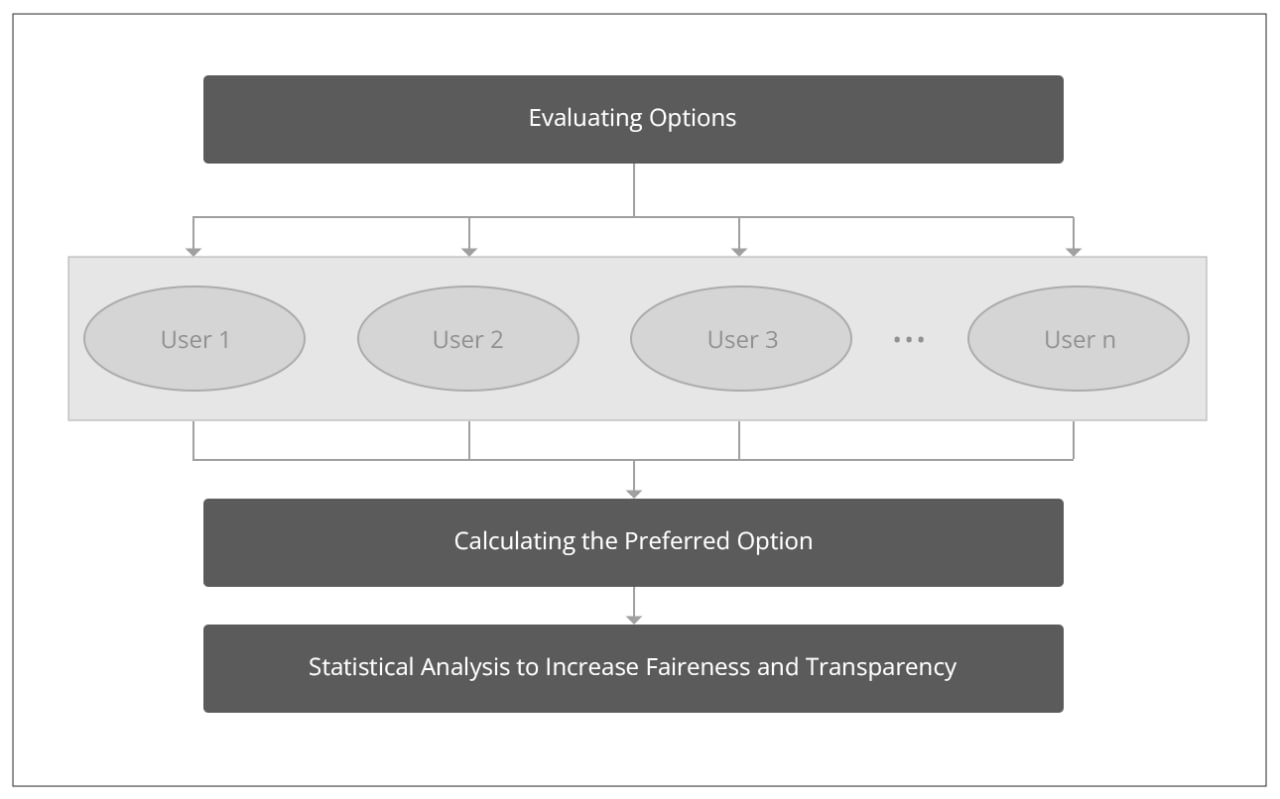}
\caption{Step 1: Voting process with the QOC DAO approach.}
\label{fig:dao_vote_step1}
\end{figure}

As depicted in Figure~\ref{fig:dao_vote_step1}, the voting interface and process guide participants through the structured evaluation of the proposal along all relevant dimensions.

Before the final decision is computed, the system applies statistical outlier detection techniques to identify potentially biased or manipulative evaluations. Votes exhibiting extreme deviations from community norms (e.g., based on standard deviation thresholds) may be excluded from the aggregation process to preserve fairness and decision quality.

Finally, a comprehensive decision report is automatically generated. This report includes a breakdown of the aggregated evaluation scores per criterion and highlights both the strengths and weaknesses as perceived by the community. For proposals that are rejected, this feedback loop enables proposers to better understand the rationale behind the decision and supports iterative refinement of future submissions.

\subsection{Step 2: Integrating AI with a Human-in-the-Loop Approach}
In the second step of the development of the QOC-based DAO governance model, the evaluation process, previously conducted manually by human stakeholders, is increasingly delegated to AI agents, enabling a semi-automated, scalable, and consistent assessment pipeline. The objective of this phase is to support decision-making through machine-driven evaluations while maintaining human oversight for the final judgment.

To begin, the system identifies the relevant stakeholder groups associated with the proposal under consideration. This may include community members, domain experts, affected users, or other interest groups. For each identified stakeholder group, one or more AI agents are instantiated. These agents are configured using structured descriptions of the stakeholder perspectives, concerns, and value orientations, either manually curated or learned from prior proposals and community interactions.

\begin{figure}[htb]
\centering
\includegraphics[width=0.7\textwidth]{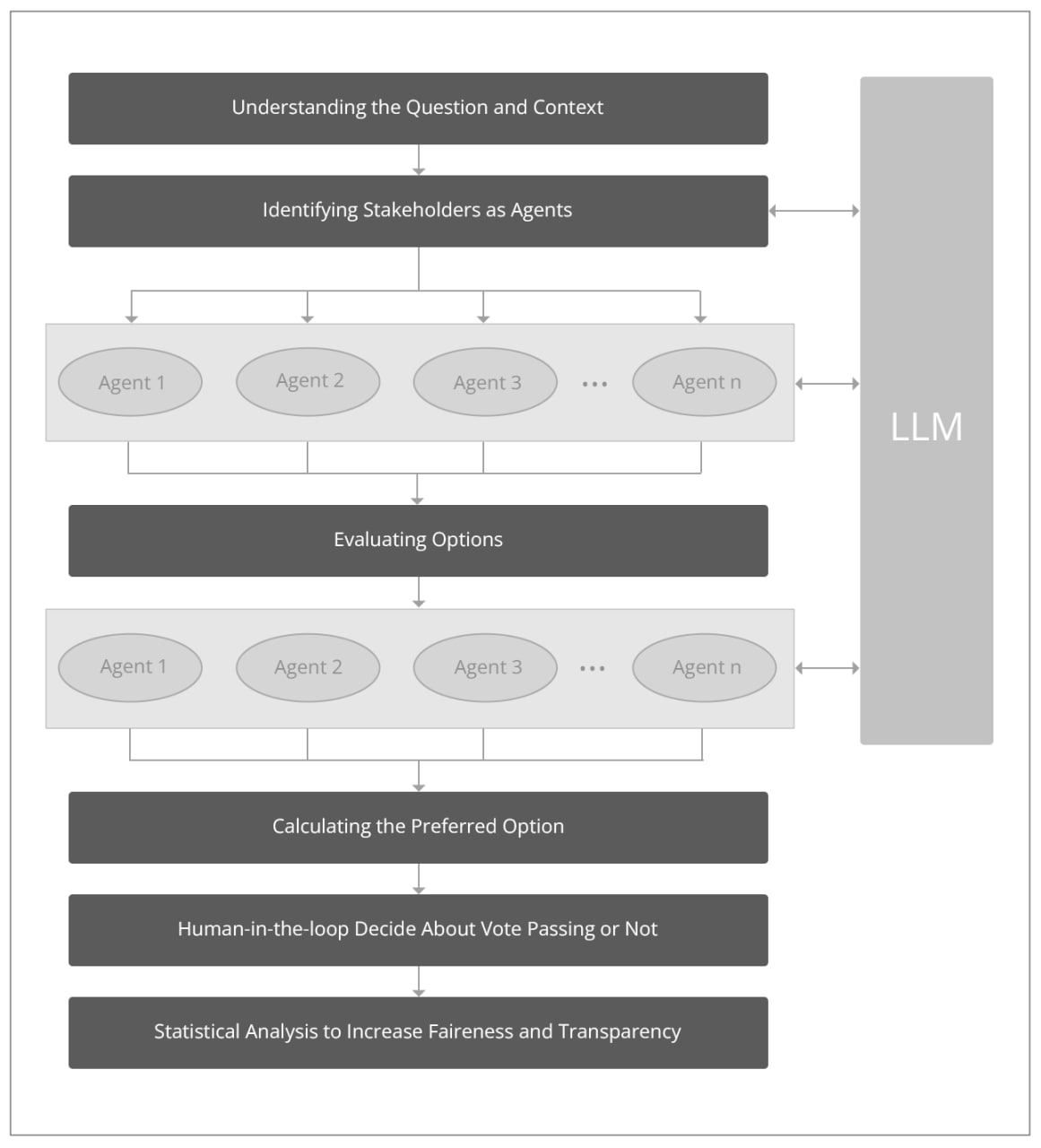}
\caption{Step 2: Voting process with AI agents and human-in-the-loop.}
\label{fig:dao_vote_step2}
\end{figure}

Once initialized, the AI agents autonomously evaluate how well the proposed initiative satisfies each of the predefined, weighted criteria. These evaluations mirror the structure of human voting in Step 1, but are now performed algorithmically based on available documentation, previous proposals, on-chain data, or external knowledge bases. The complete agent-based evaluation process is depicted in Figure \ref{fig:dao_vote_step2}.

Following the agent evaluations, the system aggregates the individual scores according to the predefined weighting schema. However, the final decision is not fully automated at this stage. Instead, the aggregated recommendation produced by the AI agents is presented to the human members of the DAO. Based on this recommendation, the human participants make the ultimate decision, either approving or rejecting the proposal. This human-in-the-loop mechanism ensures that while AI provides structured guidance and efficiency, the authority and accountability of decision-making remain with the community.

As in the first step, the system generates a comprehensive decision report. This report includes the criterion-level evaluations as derived by the AI agents, along with the final decision outcome. Such transparency not only strengthens trust in the AI-assisted process but also allows proposers to understand how their initiative was perceived algorithmically and how it aligns with the DAO’s broader goals.

This hybrid approach, combining automated analysis with human judgment, lays the foundation for future iterations of fully autonomous governance, while still preserving community involvement during critical stages of the decision-making process.

\subsection{Step 3: A Fully AI Driven DAO}
The third and final phase of the QOC-based DAO governance model represents the transition to a fully autonomous decision-making system, in which all aspects of the evaluation and resolution process are executed by AI agents, without direct human intervention. This implementation builds upon the architecture described in Step 2, including stakeholder modeling, agent instantiation, and criteria-based evaluation, but introduces a key conceptual shift: the removal of the human-in-the-loop mechanism from the decision pipeline.

As in the previous step, the process begins by identifying relevant stakeholder groups associated with the submitted proposal. Based on these stakeholder profiles, a set of corresponding AI agents is generated. Each agent is designed to simulate the perspective and evaluative tendencies of its respective stakeholder group, drawing on both static definitions and dynamic context such as previous proposal interactions, DAO documentation, and domain-specific data sources.

\begin{figure}[htb]
\centering
\includegraphics[width=0.7\textwidth]{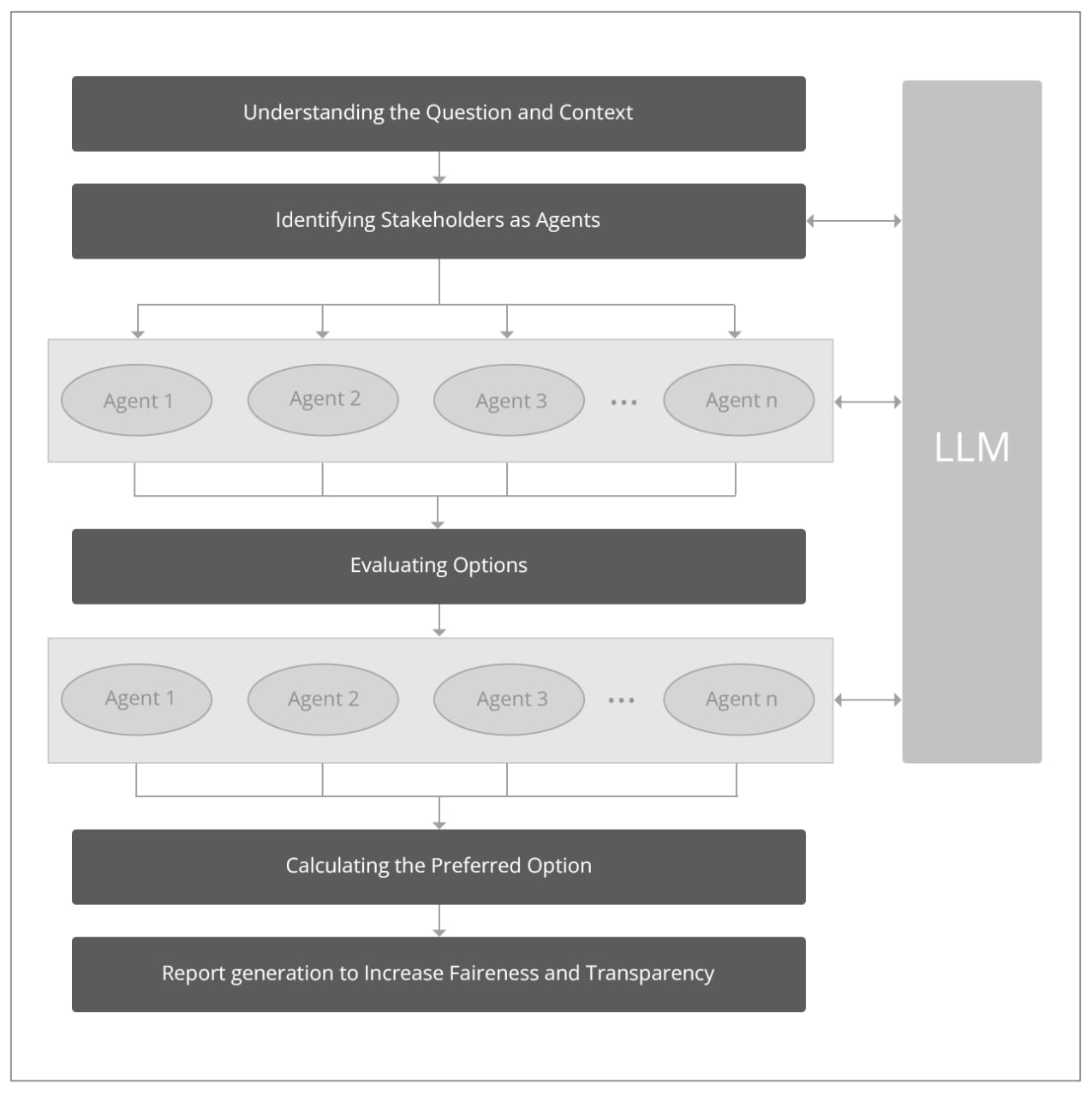}
\caption{Step 3: Fully AI-driven voting approach.}
\label{fig:dao_vote_step3}
\end{figure}

The agents then autonomously evaluate how well the proposed initiative supports each of the predefined, weighted criteria, producing a complete matrix of quantitative scores. These scores are subsequently aggregated into a final utility value per option (typically “yes” or “no”), using the weighted QOC aggregation function. The decision-making process in this step is depicted in Figure~\ref{fig:dao_vote_step3}.

The key distinction in this phase lies in the absence of human oversight: the aggregated outcome derived from the AI agents’ evaluations is directly accepted as the final decision of the DAO. There is no subsequent review, modification, or approval by human participants. This approach enables real-time, scalable, and unbiased governance, particularly valuable in contexts where rapid decision-making is essential or where human participation may be limited.

As in previous steps, the system generates a comprehensive decision report upon finalization of each vote. This report documents the aggregated evaluation scores for each criterion, provides transparency into the reasoning behind the AI-driven decision, and enables continuous feedback for applicants. In the case of proposal rejection, the report allows proposers to identify which aspects of their initiative were viewed unfavorably and to iteratively refine their application for potential resubmission.

By fully automating the governance workflow, this final implementation stage realizes the vision of self-governing, AI-enhanced DAOs, capable of making structured, explainable, and repeatable decisions at scale, without requiring human input at runtime.

\section{Evaluation}
To evaluate the proposed approach, real-world voting data were collected from an existing decentralized autonomous organization (DAO). Specifically, 102 historical voting decisions were retrieved from the PowerDAO via its public API. These decisions span a time period from December 27, 2023, to September 12, 2025, and include both accepted and rejected proposals.

To conduct the evaluation, the QOC-based decision model described above was implemented and applied to all 102 proposals using three different large language models (LLMs) from OpenAI: GPT-4-mini, GPT-5-mini, and GPT-5. Each proposal was processed independently by the respective LLM within the QOC decision framework.

Subsequently, the decisions generated by the QOC-based AI system were compared against the actual decisions made by the DAO. For each LLM, a 2×2 contingency table was constructed to visualize the agreement and disagreement between the AI-generated and DAO-recorded decisions (see Table \ref{tab:big_contingency}).
\begin{table}[h!]
\centering
\caption{Contingency table results for three LLMs compared to DAO decisions}
\label{tab:big_contingency}
\begin{tabular}{lcccccc}
\toprule
 & \multicolumn{2}{c}{\textbf{GPT-4-mini}} & \multicolumn{2}{c}{\textbf{GPT-5-mini}} & \multicolumn{2}{c}{\textbf{GPT-5}} \\

 & DAO Y & DAO N & DAO Y & DAO N & DAO Y & DAO N \\
\midrule
AI Y & 56(54.9\%) & 20(19.6\%) & 32(31.4\%) & 8(7.8\%) & 24(23.5\%) & 3(2.9\%) \\
AI N  & 14(13.7\%) & 12(11.8\%) & 38(37.3\%)  & 24(23.5\%) & 46(45.1\%) & 29(28.4\%) \\
\bottomrule
\end{tabular}
\end{table}
The rows of the table indicate the AI's decisions (AI Y = accepted, AI N = rejected), while the columns represent the corresponding DAO decisions. As shown, the degree of agreement between the AI and the DAO varies across the models: GPT-4-mini achieves the highest agreement rate (66.7\%), followed by GPT-5-mini (54.9\%) and GPT-5 (51.9\%). A notable trend can also be observed: as model quality increases, the AI becomes more conservative, more frequently rejecting proposals that were accepted by the DAO. Conversely, the number of false positives (proposals accepted by the AI but rejected by the DAO) decreases with higher-quality models.

To assess whether the differences in decisions between the DAO and the AI were statistically significant, McNemar tests \cite{mcnemar} were performed. For GPT-4-mini, the test did not reveal a statistically significant difference between AI and DAO decisions, $\chi^2(1, N = 102) = 1.06, p = .303$. However, for GPT-5-mini and GPT-5, the McNemar test indicated statistically significant differences: $\chi^2(1, N = 102) = 19.57, p = 9.72 * 10^-6$ and $\chi^2(1, N = 102) = 37.73, p = 8.11 * 10^-10$ respectively.

Although most decisions align between the AI and the DAO, the directionality of disagreement is asymmetrical: the AI tends to be more conservative, i.e., it more frequently rejects proposals that the DAO accepted than vice versa.

To further quantify the practical implications of such asymmetries, a cost function\cite{bach}\cite{ling}\cite{barnes}\cite{koch} was introduced to prioritize conservative decision-making:
\begin{center}
$c = 1 * d + 10 * a$
\end{center}
Where:
\begin{itemize}
    \item c denotes the total cost,
    \item d is the number of false negatives (proposals accepted by the DAO but rejected by the AI),
    \item a is the number of false positives (proposals rejected by the DAO but accepted by the AI).
\end{itemize}
This function reflects the assumption that false positives, i.e., approving a proposal the DAO would have rejected, are ten times more costly than false negatives. The resulting costs across the three LLMs are presented in Table \ref{tab:cost_function}:
\begin{table}[htb]
\centering
\caption{Total cost of decisions of our approach across the three used LLMs using asymmetric cost weighting}
\label{tab:cost_function}
\begin{tabular}{lccc}
\toprule
\textbf{Model} & \textbf{DAO N / AI Y} & \textbf{DAO Y / AI N} & \textbf{Total Cost} \\
\midrule
GPT-4-mini   & 20  & 14 & 214  \\
GPT-5-mini   & 8  & 38  & 118  \\
GPT-5   & 3 & 46 & 76 \\
\bottomrule
\end{tabular}
\end{table}
As expected, GPT-5, the most conservative model, yields the lowest total cost under the asymmetric penalty scheme. This suggests that despite a lower agreement rate with the DAO, its stricter approach may be preferable in contexts where false positives are particularly costly.

\section{Limitations and Future Work}
While the evaluation demonstrates promising results for automating DAO decisions using a QOC-based approach with large language models, several limitations must be acknowledged. First, the analysis was based solely on historical data from a single DAO (PowerDAO). As such, the generalizability of the findings to other DAOs or governance structures may be limited. Second, the evaluation relied on binary classification of proposals (approved vs. rejected), without accounting for nuances such as abstentions, amendments, or conditional approvals. Third, the cost function used to assess conservativeness was based on a fixed asymmetry ratio (10:1), which, while justifiable in this context, remains subjective and domain-dependent. Furthermore, the evaluation assumes static model behavior and does not account for potential temporal drift in LLM performance or governance dynamics.

In addition, although the use of AI for decision support in decentralized governance holds substantial potential for improving scalability, consistency, and efficiency, it also introduces critical questions related to trust, accountability, and control. Specifically, the delegation of evaluative power to non-human agents raises concerns about transparency, oversight, and the legitimacy of decisions made without direct human deliberation. These challenges must be addressed as part of the broader discourse on the role of artificial intelligence in decentralized systems.

Future research could explore the application of the proposed approach across multiple DAOs with varying voting cultures and proposal types. Incorporating more granular decision categories and integrating LLM uncertainty (e.g., via confidence thresholds or abstention mechanisms) may further improve alignment and trust. Moreover, future work could extend the cost model by calibrating weights empirically or dynamically based on proposal categories, risk assessments, or requested funding volumes. Finally, longitudinal studies involving real-time deployment of AI-supported governance agents would offer deeper insights into the social, ethical, and strategic implications of AI integration in decentralized decision-making processes.


\begin{thebibliography}{00}
\bibitem{intelligent_daos} Chen, J.-H., Hsu, C.-W., \& Tsai, Y.-C. (2025). Intelligent Decentralized Governance: A Case Study of KlimaDAO Decision-Making. Electronics, 14(12), 2462. https://doi.org/10.3390/electronics14122462

\bibitem{dao_design} Lustenberger, M., Küng, L., \& Spychiger, F. (2025). Designing community governance–learnings from daos. The Journal of The British Blockchain Association.

\bibitem{dao_review} Han, J., Lee, J., \& Li, T. (2025). A review of DAO governance: Recent literature and emerging trends. Journal of Corporate Finance, 102734.

\bibitem{dao_votings} Bellavitis, C., \& Momtaz, P. P. (2025). Voting governance and value creation in decentralized autonomous organizations (DAOs). Journal of Business Venturing Insights, 23, e00537.

\bibitem{blockai_medium} Verdot, C., Jansen, M. (2025). QOC , A new approach to make DAO’s more transparent and fair. https://blockai.medium.com/qoc-a-new-approach-to-make-daos-more-transparent-and-fair-24821e0c4e71

\bibitem{decision_models} Baninemeh, E., Farshidi, S., \& Jansen, S. (2023). A decision model for decentralized autonomous organization platform selection: Three industry case studies. Blockchain: Research and Applications, 4(2), 100127.

\bibitem{dao_voting_mechanisms} Tamai, S., \& Kasahara, S. (2024). DAO voting mechanism resistant to whale and collusion problems. Frontiers in Blockchain, 7, 1405516.

\bibitem{qoc} MacLean, A., Young, R., Bellotti, V., Moran, T.: Questions, options, and criteria: elements of design space analysis. Human-Comput. Interact. 6, 201–250 (1991). https://doi.org/10.1080/07370024.1991.9667168

\bibitem{mcnemar} McNemar, Q. (1947). Note on the sampling error of the difference between correlated proportions or percentages. Psychometrika, 12(2), 153-157.

\bibitem{bach} Bach, F. R., Heckerman, D., \& Horvitz, E. (2006). Considering cost asymmetry in learning classifiers. The Journal of Machine Learning Research, 7, 1713-1741.

\bibitem{ling} Ling, C., Sheng, V. (2010). Cost-Sensitive Learning and the Class Imbalance Problem, Encyclopedia of Machine Learning.

\bibitem{barnes} Barnes, B. M., \& Henn, M. A. (2023, April). Addressing misclassification costs in machine learning through asymmetric loss functions. In Metrology, Inspection, and Process Control XXXVII (Vol. 12496, pp. 352-364). SPIE.

\bibitem{koch} Koch, B., \& Imai, K. (2025). Statistical Decision Theory with Counterfactual Loss. arXiv preprint arXiv:2505.08908.
\end{thebibliography}
\end{document}